# Unidirectional Chiral Emission via Twisted Bi-layer Metasurfaces

Dmitrii Gromyko[1,2,*], Shu An[3,*], Sergey Gorelik[4], Jiahui Xu[2], Li Jun Lim[5], Henry Yit Loong Lee[3], Febiana Tjiptoharsono[3], Zhi-Kuang Tan[5], Cheng-Wei Qiu[2,+], Zhaogang Dong[3,6,+], Lin Wu,[1,7,+]

[1]Science, Mathematics, and Technology (SMT), Singapore University of Technology and Design (SUTD), 8 Somapah Road, Singapore 487372

[2]Department of Electrical and Computer Engineering, National University of Singapore, 4 Engineering Drive 3, Singapore 117583

[3]Institute of Materials Research and Engineering (IMRE), Agency for Science, Technology and Research (A*STAR), 2 Fusionopolis Way, Innovis #08-03, Singapore 138634, Republic of Singapore

[4]Singapore Institute of Food and Biotechnology Innovation, Agency for Science, Technology and Research (A*STAR), 31 Biopolis Way, #01-02 Nanos, 138669, Singapore

[5]Department of Chemistry, 3 Science Drive 3, National University of Singapore, 117543, Singapore

[6]Department of Materials Science and Engineering, National University of Singapore, 9 Engineering Drive 1, Singapore 117575

[7]Institute of High Performance Computing, A*STAR (Agency for Science, Technology, and Research), 1 Fusionopolis Way, #16-16 Connexis, Singapore 138632.

[*]These authors contributed equally to this work
[+]Corresponding authors: chengwei.qiu@nus.edu.sg; dongz@imre.a-star.edu.sg; lin_wu@sutd.edu.sg

**Abstract**

Controlling and channelling light emissions from unpolarized quantum dots into specific directions with chiral polarization remains a key challenge in modern photonics. Stacked metasurface designs offer a potential compact solution for chirality and directionality engineering. However, experimental observations of directional chiral radiation from resonant metasurfaces with quantum emitters remain obscure. In this paper, we present experimental observations of unidirectional chiral emission from a twisted bi-layer metasurface via multi-dimensional control, including twist angle, interlayer distance, and lateral displacement between the top and bottom layers, as enabled by doublet alignment lithography (DAL). First, maintaining alignment, the metasurface demonstrates a resonant intrinsic optical chirality with near-unity circular dichroism of 0.94 and reflectance difference of 74%, where a high circular dichroism greater than 0.9 persists across a wide range of angles from -11 to 11 degrees. Second, engineered lateral displacement induces a unidirectional chiral resonance, resulting in unidirectional chiral emission from the quantum dots deposited onto the metasurface. Our bi-layer metasurfaces offer a universal compact platform for efficient radiation manipulation over a wide angular range, promising potential applications in miniaturized lasers, grating couplers, and chiral nanoantennas.

Collimation and directional radiation from quantum emitters pose challenges across various optical applications, including sensing, optical communication[1], and quantum optics[2-4]. Traditional solutions involve bulky optical components like lenses, beam splitters, and mirrors, while compact lasers[5,6] and integrated photonic technologies[7] demand efficient miniaturized optical devices capable of emission amplification and radiation channeling. Various methods have been explored for directional emission control, including nano-antennas[8-11], phase-gradient metasurfaces[12], hybrid plasmon-emitter coupled metasurfaces[13,14], chiral photonic-crystal waveguides[15], and photonic lattices under vortex excitation[16]. Unfortunately, these approaches often require precise emitter positioning, emitter chirality, or complex excitation techniques.

Dielectric metasurfaces with unidirectional resonant modes[17-20], emitting solely to one side, show promise due to their extended structure and near-field emission enhancement. However, experimental demonstrations are rare, and control over emission polarization is lacking. Meanwhile, various resonant nanostructures with strong chiral effects *at normal light incidence* have been explored, including multi-height metasurfaces[21-24], flat metasurfaces[25-30], slant-geometry metasurfaces[31,32], stacked and twisted bi-layer nanostructures[33-37], and chiral nanocavities[38,39], yet addressing emission directionality remains unexplored. Efforts have also been made to achieve absolute chirality *at oblique light incidence* using bound-in-continuum (BIC) resonances[40-42] or precise engineering of synthetic valleys in the folded Brillouin zone[43]. Yet, these designs often retain certain mirror symmetries, resulting in symmetric polarization and intensity patterns of emission. In most of these cases, chiral resonances in metasurfaces evolve from the achiral modes found in flat metasurfaces and, thus, are considered indivisible entities, missing one or several degrees of design freedom. Stacked designs with vertical and lateral offsets naturally introduce a phase shift into the electromagnetic wave scattering problem, allowing for chirality[23] and directionality engineering[44]. However, experimental observations of directional chiral radiation from resonant metasurfaces with near-field-coupled emitters remain obscure, likely due to the challenge of achieving multidimensional precise control.

In this paper, we design and fabricate twisted bi-layer metasurfaces and demonstrate the chirality and radiation directionality of resonant modes. As illustrated in Fig. 1a, the design comprises two layers of metasurfaces, where a square array (pitch $p$) of silicon (Si) discs (radius $R$ and thickness $h$) with rectangular notches (depth $d$ and width $w$) are embedded in fused silica. This design has three degrees of freedom for full control of the optical response, as depicted in Fig. 1b: (i) the notches cut at angles $\mp\theta$ with respect to the $y$-axis in the top and bottom discs; (ii) the top and bottom metasurfaces separated by the distance $H$; (iii) the centers of the top and bottom discs shifted in lateral directions by $\Delta X$ and $\Delta Y$. We will showcase multiple configurations of bi-layer metasurfaces with broad potential for light manipulation: the red, blue, and black squares in Fig. 1c-e indicate configurations A1/A2 (aligned, no lateral shift), D1, and D2 (displaced), respectively. In the near-field coupling regime with small interlayer separation (A1), the bi-layer twisted metasurface exhibits a single resonance of wide-angle intrinsic chirality. At the intermediate coupling regime with enlarged interlayer separation (A2), two closely spaced intrinsically chiral resonances suitable for chirality switching emerge. The introduction of lateral displacement between the constituent layers (D1 and D2) dramatically modifies the radiation properties of the resonant mode in reciprocal space, revealing a unidirectional chiral resonance. Integration of this metasurface with perovskite quantum dots (QDs) results in a unique emission pattern in the top half-space, with resonant circularly polarized (CP) light emitted along positive *x* and *y* directions while suppressing radiations in the negative *x* and *y* directions (illustrated in Fig. 1). We term this effect "unidirectional chiral emission". This work suggests the potential for full multi-dimensional control of polarization response, enabling reconfigurable metasurface design via its integration with nano-electro-mechanical systems[45] or via scaling to the terahertz frequencies[46].

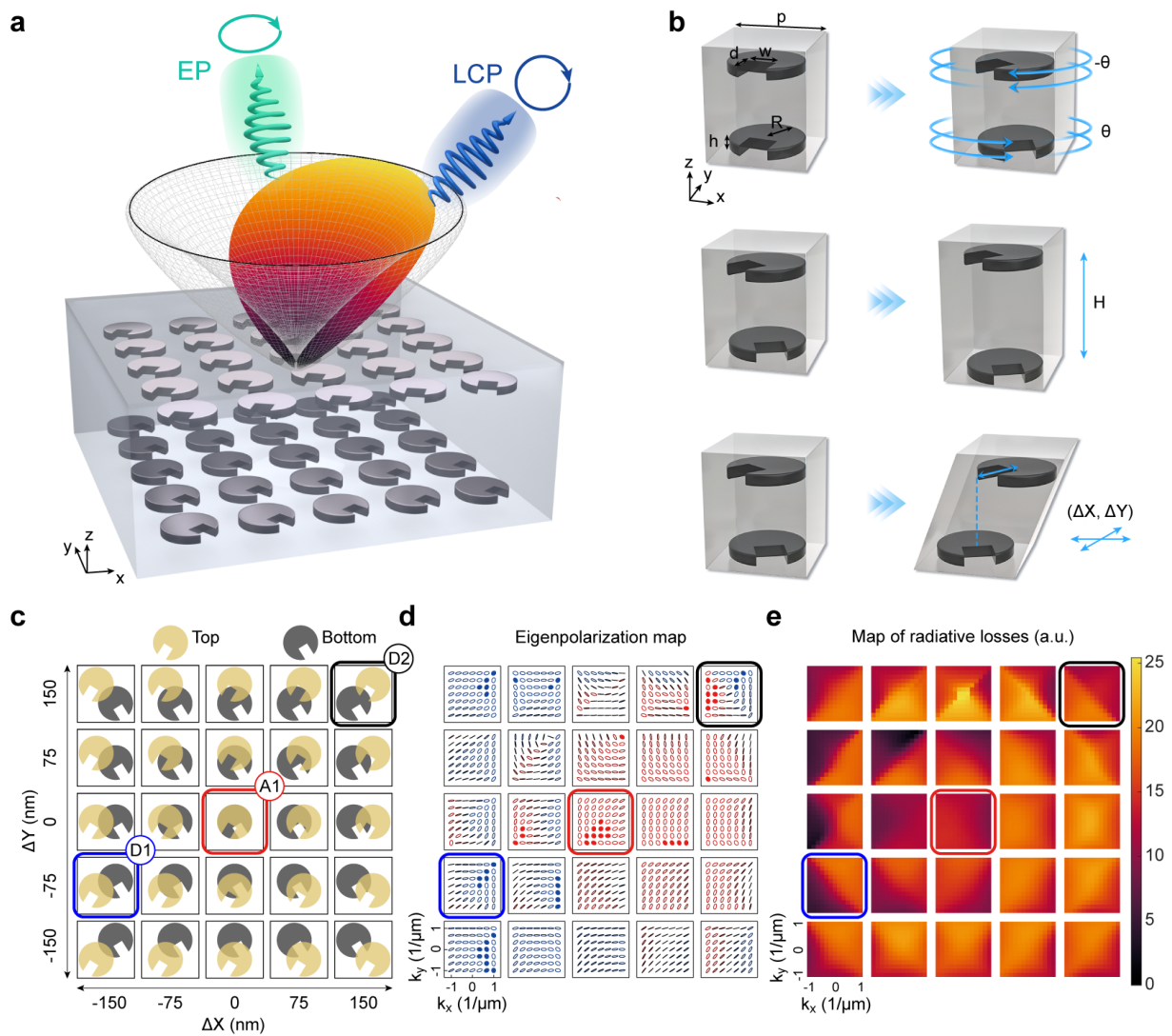

**Fig. 1 | Multi-dimensional control of radiation directionality and polarization patterns from twisted bi-layer metasurface. a,** Twisted bi-layer metasurface and a schematic illustration of the unidirectional chiral emission pattern in the top half-space as per experimentally measured emission in Fig. 4. **b,** Three designing parameters applied to the structure unit cell: twist angle $\theta$ between the *y*-axis and notches of the top and bottom discs, interlayer distance $H$, and lateral displacement ($\Delta X, \Delta Y$). The unit cell is characterized by pitch $p$, disc radius $R$, and thickness $h$, depth $d$ and width $w$ of rectangular notches. **c,** Illustrations of bi-layer metasurfaces with possible lateral shifts ranging from -150 nm to 150 nm in both directions. **d, e,** Corresponding maps of eigenpolarization and radiative losses for the anti-symmetric resonant mode in the reciprocal space.

# Results

**Designing twisted bi-layer metasurfaces**

We commence the analysis by considering metasurfaces comprising discs without notches: each constituent single-layer metasurfaces have a four-fold in-plane rotational symmetry ($D_{4h}$), and all non-degenerate resonances below the diffraction opening are BIC at Γ-point (normal light incidence). The introduction of notches breaks the rotational symmetry and allows polarization-selective coupling to the radiation continuum (*i.e.*, the notch size serves as the asymmetry parameter[47]), which can be conveniently described using an effective dipole model (SI-1). Owing to the high symmetry of the original near-field distribution (SI-2), the lowest transverse electric (TE) resonance of a single-layer metasurface with a notch angle $\theta$ selectively interacts with waves polarized as $\mathbf{e} = (\cos\theta, \sin\theta)$.

Due to interlayer coupling, the resonances of the bi-layer metasurface are hybrids of the modes from the two constituent metasurfaces. The dependence of the resonant frequencies on the interlayer distance $H$ is shown in Fig. 2a. When the constituent metasurfaces are well-aligned ($\Delta X = \Delta Y = 0$), the bi-layer metasurface exhibits a two-fold symmetry of rotation around the *y*-axis; hence, all the hybrid modes transform like irreducible representations of the $C_2$ group. As shown in Fig. 2b, the two lowest-energy modes are hybrids of the TE modes in the constituent metasurfaces, categorized as *symmetric* and *anti-symmetric* with respect to $C_2$ "flipping" symmetry.

Tunning of $\theta$ and $H$ is the key to achieving the absolute intrinsic chirality of these modes. To evaluate the far-field response of the bi-layer metasurface, we approximate the electric fields of the resonant modes in constituent metasurfaces with dipoles, $\mathbf{D}_1^{\text{self}} = (\cos\theta, -\sin\theta)$, $\mathbf{D}_2^{\text{self}} = \pm(\cos\theta, \sin\theta)$, where $\pm$ sign stands for the anti-symmetric and symmetric with mode, respectively. During coupling, the excited modes emit far-field radiation that propagates through the intermediate layer and induces nonresonant electric fields in the counterpart metasurfaces with coupling coefficient $r$ being the background reflection coefficient of a single-layer metasurface (details of the coupling model are provided in SI-3,4). These induced fields are approximated by the dipoles $\mathbf{D}_{1,2}^{\text{ind}}$. The coupled dipoles appear as a sum of the resonant and nonresonant terms:

$$\mathbf{D}_1^{\text{ind}} = re^{-ik_0 nH}\mathbf{D}_2^{\text{self}}, \qquad \mathbf{D}_2^{\text{ind}} = re^{-ik_0 nH}\mathbf{D}_1^{\text{self}},$$

$$\mathbf{D}_n = \mathbf{D}_n^{\text{self}} + \mathbf{D}_n^{\text{ind}}, \qquad n = 1,2. \tag{1}$$

Due to the nonzero imaginary part of the coupling-induced fields $\mathbf{D}_n^{\text{ind}}$, light scattered by the top and bottom metasurfaces generally has an elliptical rather than linear polarization.

Under RCP and LCP waves with electric fields $\mathbf{e}^{\text{R,L}} = (1, \pm i)/\sqrt{2}$ normally incident on the bi-layer metasurface from the top, a hybrid mode is excited with amplitudes:

$$\boldsymbol{\alpha}^{R,L} = \mathbf{e}^{R,L}\mathbf{D}_1 + e^{-ik_0 n(H+h)} \mathbf{e}^{R,L}\mathbf{D}_2. \tag{2}$$

The normalized third Stokes parameter or the resonance far-field polarization, which we call the resonance chirality (RC), can be evaluated as:

$$S_3 = \frac{|\boldsymbol{\alpha}^R|^2 - |\boldsymbol{\alpha}^L|^2}{|\boldsymbol{\alpha}^R|^2 + |\boldsymbol{\alpha}^L|^2}. \tag{3}$$

Using our coupled dipole model, we calculate RC for the bi-layer metasurfaces with varied $\theta$ and $H$ (Fig. 2c,d, left panels), demonstrating complete agreement with the full-wave simulations (Fig. 2c,d, right panels).

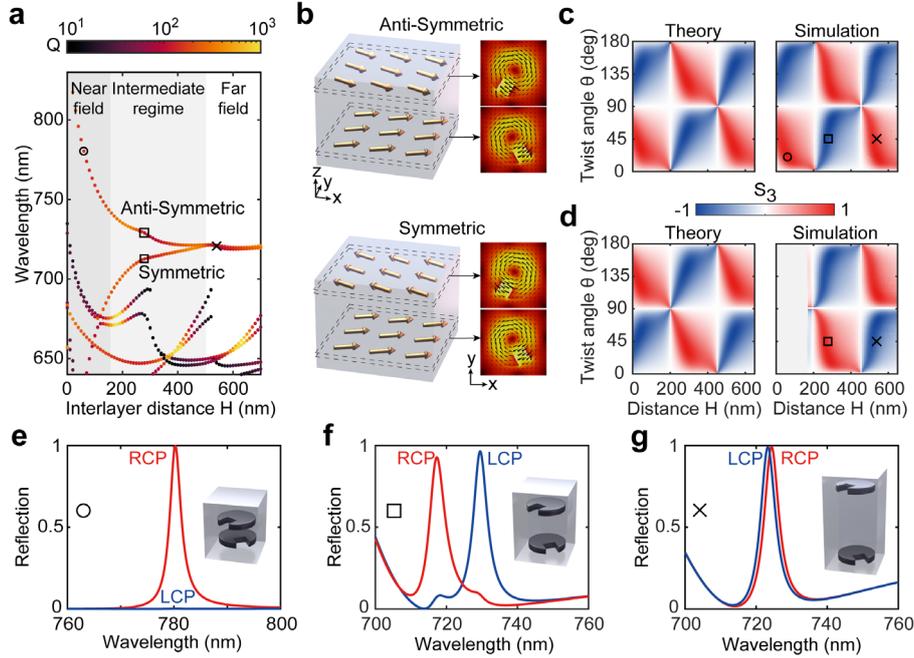

**Fig. 2 | Design principles for the twisted bi-layer metasurfaces**. **a,** Resonant energies of the coupled modes in the bi-layer metasurface in dependence on the interlayer distance. **b,** Dipole representation of the anti-symmetric and symmetric coupled modes of the bi-layer metasurface. **c, d,** Resonance chirality maps of the anti-symmetric and symmetric hybrid modes in twisted bi-layer metasurfaces with varied twist angle $\theta$ and interlayer distance $H$. Grey area in **d** indicates the range of interlayer distances at which the symmetric mode strongly hybridizes with the higher-order modes. **e, f, g,** Reflection spectra of the bi-layer metasurfaces that operate in the near-field, intermediate, and far-field coupling regimes, respectively. The circle, square, and cross signs correspond to structure parameters ($H$=50 nm, $\theta$=20º), ($H$=280 nm, $\theta$=45º), and ($H$=530 nm, $\theta$=45º) shown in **a**, **c**, and **d**.

As indicated in Fig. 2a, the mode coupling can be divided into three regimes: near-field, intermediate, and far-field. In the near-field regime ($H$ ~ notch size), the symmetric and anti-symmetric modes are well separated spectrally, where only the anti-symmetric mode with the lowest energy is available, and the symmetric mode couples to higher order resonances and even radiates into the first diffraction order. According to the chirality map (see Fig. 2c), the optimized near-field design ($H = 50$ nm, $\theta = 20^0$) exhibits absolute intrinsic chirality in Fig. 2e. Meanwhile, strong near-field coupling ensures that even the modes of moderately dissimilar metasurfaces (*e.g.*, different dielectric environments) can hybridize into chiral modes (SI-5). Thus, in the near-field regime, we can design an "open structure" with only the bottom discs fully embedded in silica and the top discs in the air. This enables a versatile chiral metasurface platform to be integrated with quantum emitters. In the intermediate regime ($H \sim$ disc size), the energy splitting, albeit weaker, is still governed by the near fields that resemble those of the unperturbed BIC mode. However, in this regime, *e.g.*, ($H = 280$ nm, $\theta = 45^o$), both symmetric and anti-symmetric quasi-BIC modes are accessible and exhibit high chirality while coupling to waves with opposite circular polarizations (Fig. 2f). When $H$ exceeds the metasurface period, purely far-field coupling determines the properties of the coupled resonances. The energy splitting becomes insufficient, and two chiral resonances spectrally overlap (see Fig. 2a,g), no matter how $\theta$ changes (SI-6).

**Intrinsic wide-angle chirality**

The designed metasurface of configuration A1 (open structure, $\theta = 32^o, H = 60$ nm, $\Delta X = \Delta Y = 0$ nm, other parameters listed in Methods) operated in the **near-field regime** is fabricated by a DAL technique (see details in the Methods). The scanning electron microscope (SEM) image shown in Fig. 3a confirms the perfect alignment of the top and bottom layers of metasurfaces. The corresponding map of the calculated anti-symmetric resonance eigenpolarization in the reciprocal space is shown in Fig. 3b, indicating maximum intrinsic chirality. Moreover, the resonance is primarily RCP-polarized at all oblique incidence angles. Fig. 3c and Fig. 3d demonstrate the measured angle-resolved reflection spectra for the structure under LCP and RCP illumination, which we denote $R^{\text{LCP}}$ and $R^{\text{RCP}}$. Examining the spectra under normal incidence (Fig. 3e), we calculate the maximum circular dichroism (CD), which is defined as $CD = (R^{\text{RCP}} - R^{\text{LCP}})/(R^{\text{RCP}} + R^{\text{LCP}})$ to be 0.94 at resonance and the maximum reflectance difference $\Delta R$ of 74% (experiment). Due to the relatively large notch size, wide-angle intrinsic chirality with a high CD larger than 0.9 is observed at angles ranging from -11º to 11º, as shown in Fig. 3f.

To observe chiral modes in the **intermediate** regime, we investigate the metasurface of configuration A2 ($\theta = 18^o$, $H = 280$ nm, $\Delta X = \Delta Y = 0$ nm, other parameters listed in Methods) for maximized chirality of the symmetric mode. Configuration A2 is fully embedded into fused silica, with a 500-nm-thick fused silica layer above the top metasurface. By designing a smaller notch size, the resonant modes have a greater radiative Q-factor than the modes of configuration A1. This allows one to spectrally separate the two resonant features associated with the symmetric and anti-symmetric modes. Fig. 3g and Fig. 3h illustrate the measured LCP and RCP reflection spectra, respectively. Two sharp resonant modes are observed; however, the magnitude of resonant reflection is not as high as that for configuration A1. This should be attributed to small radiative losses of the resonance and the presence of nonradiative losses, which are inevitable even in all-dielectric structures due to the finite size of the sample and scattering on fabrication irregularities. Nevertheless, configuration A2 demonstrates a chiral response at a normal incidence of illumination associated with the symmetric mode (Fig. 3i), which accords well with the predictions of the dipole model.

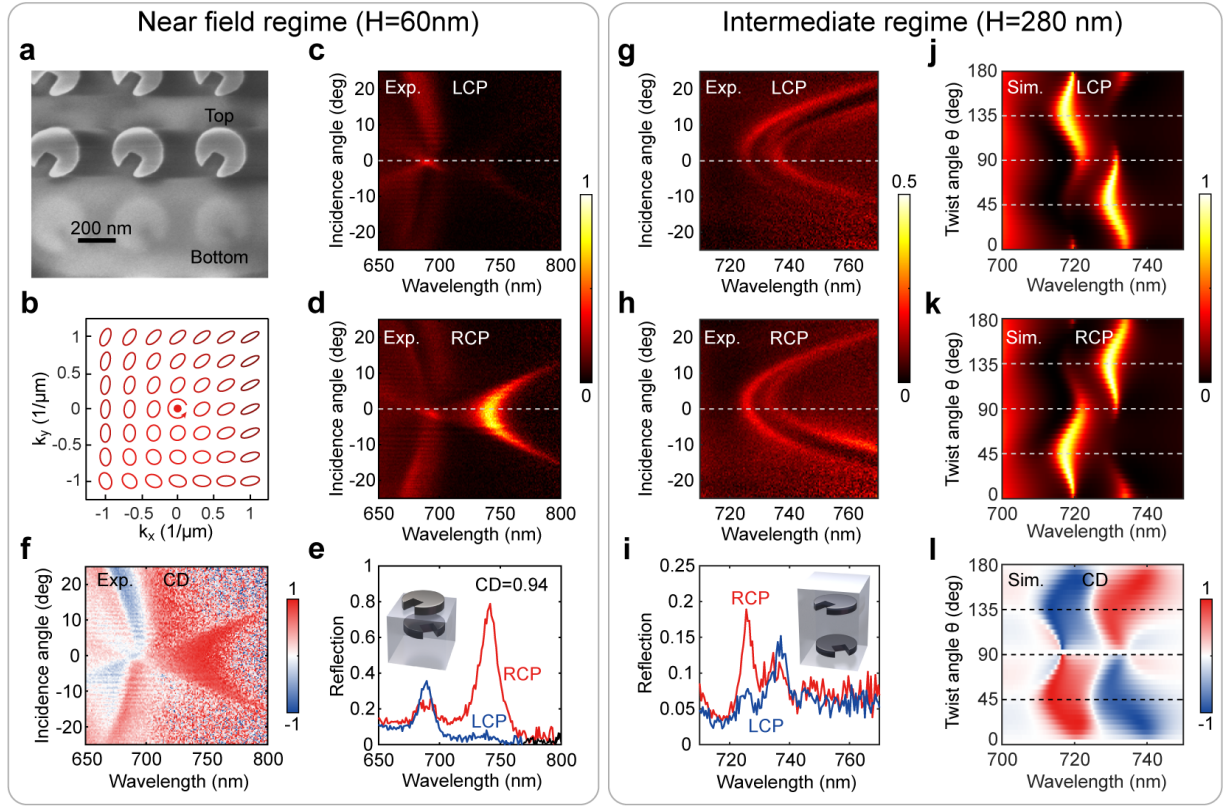

**Fig. 3 | Twisted bi-layer metasurfaces operating in near-field and intermediate coupling regimes**. **a,** SEM image of the bi-layer twisted metasurface with the near-field coupling. The total area of the top metasurface is slightly smaller than the area of the bottom metasurface, thus allowing us to visualize both layers near the edge of the top metasurface. **b,** Calculated eigenpolarization map of the anti-symmetric resonance of the near-field coupling design. **c, d,** Measured angle-resolved reflection spectra under LCP and RCP illumination. **e,** Measured reflection spectra at normal incidence corresponding to the dashed lines in **c**, **d**. **f,** Angle-resolved circular dichroism (CD) heatmap. **g, h,** Measured LCP and RCP angle-resolved-reflection spectra of bi-layer twisted metasurface in the intermediate coupling regime. **i,** Corresponding reflection spectra at normal incidence. **j, k,** Simulated LCP and RCP reflection spectra at normal incidence of the bi-layer metasurface with variable twist angle operating in the intermediate coupling regime. **l,** CD spectra calculated based on **j** and **k**.

The two sharp resonant modes in the intermediate regime can be potentially exploited for chirality switching. We show simulated reflection coefficients with variable $\theta$ under LCP and RCP illumination in Fig. 3j and Fig. 3k. Considering a bi-layer metasurface with twist angle $\theta=45º$, symmetric and anti-symmetric resonances predominantly couple to RCP and LCP incoming waves, respectively. At $\theta=90º$ (notches at opposite sides), the chirality vanishes as the effective dipoles of such modes are either co- or counter-directional. At $\theta=135º$, the modes become chiral again, albeit the chirality of the modes switches (Fig. 3l). Such changes in the angle $\theta$ can be achieved through rotation of the top and bottom metasurfaces with respect to each other. Although, in general, the unit cell size of two stacked and twisted metasurfaces forming a Moiré pattern[48] differs from the original unit cell size of a single metasurface, relative rotations by 2Δθ=90° preserve the unit cell, thus avoiding the undesired diffraction effects. In this sense, a square or hexagonal unit cell with a notched disc is preferable for bi-layer chiral structures. We envision this kind of system with an ability for relative Moiré-like rotation of stacked quasi-BIC layers as a prospective candidate for an active chirality switching device.

**Unidirectional chirality engineering by lateral displacement**

One common feature of all stacked structures, namely, layer lateral alignment, is another important parameter that allows the engineering of both the polarization and directionality characteristics of the bi-layer metasurface. The proposed coupling model assumes precise alignment of the top and bottom discs. Under this condition, the unperturbed bi-layer metasurface without notches exhibits 4-fold rotational symmetry with respect to the vertical $z$-axis, ensuring that both symmetric and anti-symmetric modes are BIC. However, the introduction of lateral displacement breaks the symmetry, rendering the modes radiative even without notches. Thus, two types of symmetry breaking-etched notches and lateral displacement-determine the polarization selection rules, providing additional design flexibility (Fig. 1c). To systematically investigate the effect of lateral displacement, we calculate eigenpolarization maps and maps of radiative losses in the top half-space for the anti-symmetric mode of bi-layer metasurfaces with varied displacement in Fig. 1d,e (an extended map with finer resolution if presented in SI-7). All other structure parameters remain consistent with A1.

Metasurfaces featuring selected displacements can exhibit significant asymmetry in radiative losses in reciprocal space. For example, configuration D1, highlighted in the blue square in Fig. 1c with a displacement of $ΔX = -150$ nm, $ΔY = -75$ nm, is expected to resonantly radiate LCP waves along the positive $x$ and $x = y$ directions, as well as elliptically polarized waves along the positive $y$ direction (refer to Fig. 1d,e). Conversely, resonance radiation along the negative $x$, $y$, and $x = y$ directions is notably suppressed. Similar maps illustrating resonance radiation properties in the bottom half-space are provided in SI-7, showing that total radiative losses do not completely vanish but are redistributed between upward and downward propagating radiation, resulting in unidirectionality across a broad area of reciprocal space.

In contrast, configuration D2 with a different displacement of $ΔX = +150$ nm, $ΔY = +150$ nm (indicated by the black square in Fig. 1c-e) does not demonstrate significant radiation asymmetry in the top half-space but exhibits pronounced extrinsic chirality. To underscore the various possibilities of resonance polarization control, we analyze experimental angle-resolved resonant reflection spectra for configuration D2 in SI-8, registering CD values up to 0.85.

**Unidirectional chiral emission**

The designed twisted bi-layer metasurface D1 can be exploited as a highly asymmetric router of light to achieve unidirectional chiral emission when it is integrated with active materials, such as perovskite QD, as illustrated in Fig. 1a. A thin layer of $FAPbI_3$ QDs is deposited on top of DAL-fabricated metasurface D1 (Fig. 4a) using the drop-cast technique[49]. The QDs (Fig. 4b) are excited using a 532 nm continuous wave (CW) pump laser, and the photoluminescence signal is harvested in the top half-space in the 650-850 nm wavelength range. Fig. 4c shows the cross-section of the bi-layer metasurface in the *x-z* plane and the resonant electric field profile for the anti-symmetric mode at $(k_x, k_y) = \pm(1.05, 1.05)$ 1/ μm. The patent unidirectionality at these opposite points of the reciprocal space also leads to pronounced emission asymmetry at opposite incidence angles. To visualize the radiation intensity and polarization patterns, we simulate emission at 770 nm wavelength from a point dipole source located at the center of the top disc notch (see Fig. 4d). As expected from the radiative losses and polarization maps in Fig. 1, emission in the top half-space is suppressed along the negative *x* and *y* directions because of full redirection into the bottom half-space (see SI-9), while along the positive *x* direction, the emission is LCP. The additional emission intensity peaks along the diagonal directions are associated with the higher-frequency symmetric resonant mode.

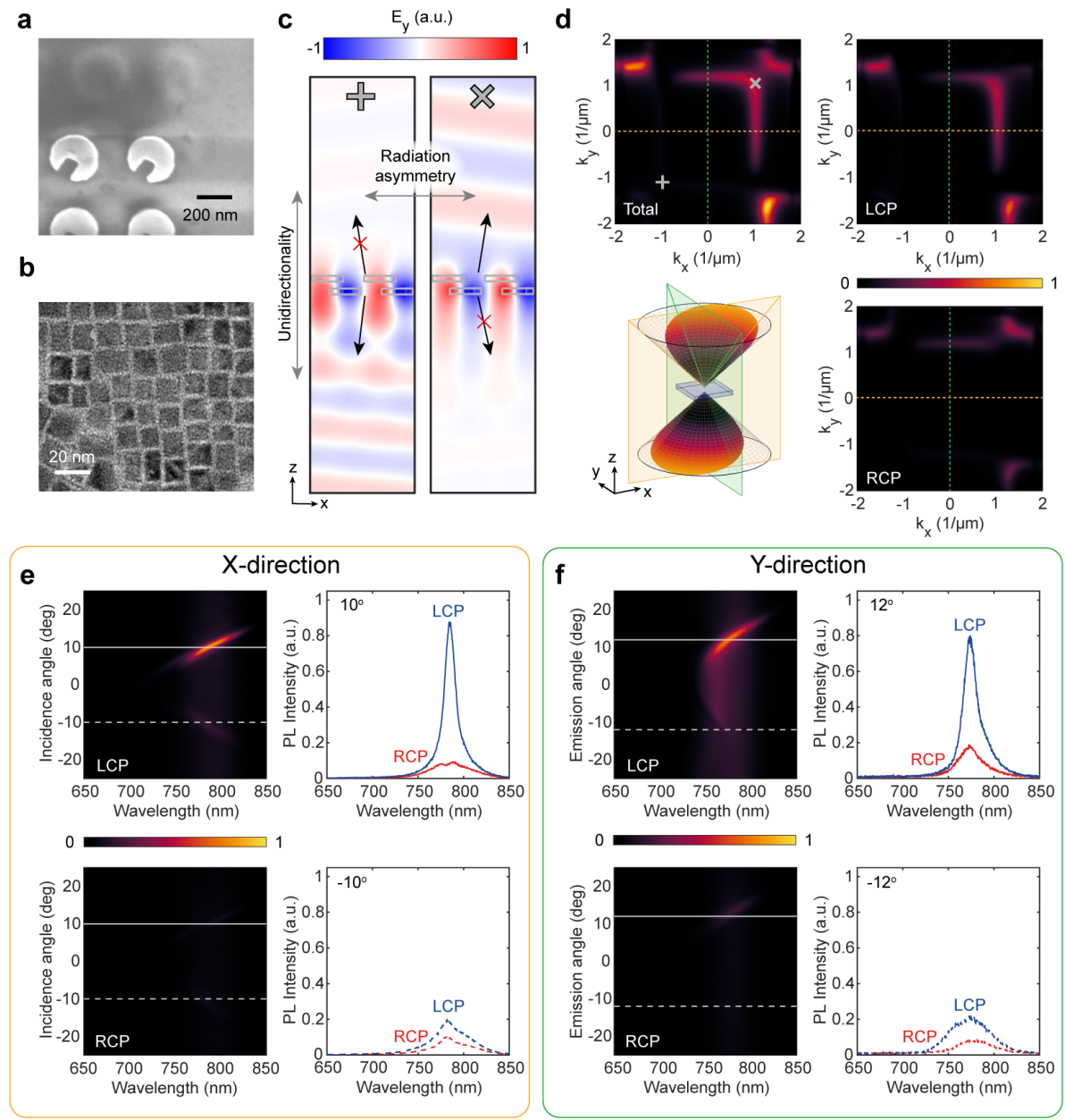

**Fig. 4 | Unidirectional chiral emission. a,** SEM image of the laterally displaced bi-layer twisted metasurface. **b,** Transmission electron microscope (TEM) image of the perovskite QDs. **c,** Resonant

electric field profiles at two opposite points of reciprocal space $(k_x, k_y) = \pm(1.05, 1.05)$ 1/μm. **d,** Calculated emission intensity at wavelength 770 nm. **e,** Left column: Measured angle-resolved photoluminescence spectra in the *x-z* plane (LCP and RCP components). Right column: Emission intensities at $\pm 10^o$. **f,** Left column: Measured angle-resolved photoluminescence spectra in the *y-z* plane (LCP and RCP components). Right column: Emission intensities at $\pm 12^o$. The overall radiation pattern in the top half-space has been systematically illustrated in Fig. 1a.

We measure the emission of QDs along the *x* and *y* directions, represented by the orange and green planes in Fig. 4d, respectively. Experimentally measured angle-resolved and polarization-resolved emission intensities along the *x* direction are depicted in Fig. 4e. Positive emission angles ($k_x > 0$) exhibit a strong resonant LCP signal, while negative angles show suppressed resonance radiation, resulting in a nearly background emission profile. These results align with numerical emission simulations (see SI-9). Emission spectra measured along the *y* direction, shown in Fig. 4f, also demonstrate similar radiation asymmetry: the intensity of elliptically polarized resonant emission, with a dominant LCP component at positive emission angles, significantly exceeds that at negative angles. Both experimental and simulated angle-resolved emission spectra showcase asymmetric chiral emission patterns due to the property of unidirectionality, a feature not achieved in any previous designs for extrinsic chirality[40-43]. These observations corroborate that the radiation unidirectionality in the bi-layer metasurface is not a special resonance feature at a single point in the reciprocal space but rather a broad-angle effect.

Additionally, we measure the emission spectra from D2 integrated with the QDs (see SI-8). While the emission intensity is symmetric with respect to the emission angle, the polarization of the emission associated with the resonant mode changes drastically from RCP at $k_y = 0, k_x < 0$ to LCP-dominated elliptical at $k_y = 0, k_x < 0$.

## Discussion

In summary, we have designed and demonstrated twisted bi-layer metasurfaces made of notched dielectric discs, showcasing their potential as a promising platform for simultaneous control over light polarization and propagation direction. First, precise tailoring of interlayer distance and the angle between the notches of the top and bottom discs introduces structure asymmetry that enables tunable resonant modes selective to specific wave polarizations. For structures with subwavelength interlayer distances and optimized chirality, measured CD values at normal incidence reach 0.94, with reflection differences of up to 74%. Second, by increasing the interlayer distance, we experimentally demonstrate the emergence of two hybrid quasi-BIC modes and propose a mechanism of active chirality switching for these modes. Third, the introduction of lateral displacement between the top and bottom layers allows manipulation of resonance directionality and eigenpolarization patterns in reciprocal space, revealing a metasurface with unidirectional resonance and strong radiation asymmetry. Finally, we demonstrate unidirectional circularly polarized emission from the bilayer twisted metasurface integrated with perovskite quantum dots. We anticipate that our design can be brought to lower operational frequencies and then readily integrated with nano-electro-mechanical systems to achieve *reconfigurable* chiral metasurface systems capable of actively manipulating emission angle, wavelength, and polarization.

## Methods

**Simulation** All calculations were conducted using COMSOL Multiphysics implementation of the finite element method. Periodic boundary conditions with Floquet periodicity in $x$ and $y$ directors were used with two ports on the top and bottom of the structure ($z$-direction). Eight additional diffraction channels (diffraction orders $(\pm 1,0)$ and $(0,\pm 1)$ with in-plane and out-of-plane polarization) were turned on in the bottom port to simulate diffraction into the substrate. Optical spectra of the structures were calculated using the wavelength domain study with excitation from the top port, while the resonant properties of the structure were calculated using eigenfrequency study. Simulations of the resonant modes of the metasurfaces were conducted using the COMSOL eigenfrequency solver. COMSOL build-in integration functions were applied to the simulated resonant fields to calculate the effective dipoles. Additional calculations based on the analytical coupling model and resonant dipoles post-processing were done in MATLAB.

To simulate the resonance-assisted photoluminescence emission, the top and bottom ports were replaced with scattering boundary conditions. A point emitter is characterized by a dipole moment $d = (\cos\theta, -\sin\theta)/(\omega - \omega_r - i\gamma)$ was put in the center of the top disc notch to model a perovskite QD. The emitter was characterized by resonant frequency $\omega_r = 384.6$ THz (which corresponds to a wavelength $\lambda = 780$ nm) and damping factor $\gamma = 13$ THz, which ensured a relatively broad range of emission wavelengths and a good approximation of the background emission dependence on the wavelength observed in the experiment.

All simulations related to the theory of intrinsically chiral resonances in bi-layer metasurfaces with alignment (Fig. 2) were conducted using the following parameters: pitch $p = 400$ nm, disc radius $R = 145$ nm, thickness $h = 50$ nm, depth $d = 100$ nm and width $w = 80$ nm of rectangular notches. The complex refractive index of silicon was obtained from ellipsometry measurements, the refractive index of the embedding silicon dioxide is 1.45. For better simulation-to-experiment comparison, supporting simulations for experimentally fabricated samples were conducted with a silicon dioxide refractive index of 1.52. Configuration A1 had parameters $p = 450$ nm, $R = 135$ nm, $h = 50$ nm, $d = 115$ nm, $w = 80$ nm, $\theta = 32°, H = 60$ nm, $\Delta X = \Delta Y = 0$ nm, top discs were in the air (open structure). Configuration A2 had parameters $p = 400$ nm, $R = 145$ nm, $h = 50$ nm, $d = 100$ nm, $w = 80$ nm, $\theta = 18°, H = 280$ nm, $\Delta X = \Delta Y = 0$ nm, the structure was fully embedded in silicon dioxide. Configurations D1 and D2 had the same structural parameters as structure A1 but with $\Delta X = -150$ nm, $\Delta Y = -75$ nm, and $\Delta X = 180$ nm, $\Delta Y = 120$ nm, respectively.

**Fabrication** The fabrication was done using the doublet alignment lithography (DAL) technique (see SI-10). The bi-layer metasurfaces were fabricated layer by layer with a nanofabrication process based on electron beam lithography (EBL), lift-off, etching, and alignment. First, an amorphous Si (a-Si) layer with a thickness of 50 nm was grown on a thoroughly cleaning glass substrate (length and width of 26 mm and 20 mm) using inductively coupled plasma-enhanced chemical vapor deposition (ICP CVD, Oxford Instrument Plasmalab System. 100) (Fig. S13a, S13b). To prepare for the well-aligned bi-layer metasurface, a pair of gold (Au) alignment markers with a distance of 16 mm was patterned by EBL, followed by Au evaporation and lift-off process (Fig. S13c). The bottom Si disk is made by patterning and over-etching (Fig. S13d). The notch pattern was well aligned using the Au alignment marker during the first EBL process to avoid the misalignment between the top and bottom disks. The patterned a-Si sample was then etched using a chlorine ($Cl_2$)-based inductively-coupled-plasma reactive-ion etching (ICP RIE, Oxford Instruments Plasmalab System 100) with an ICP power of 300 watts, an RF power of 100 watts under a pressure of 5 mTor and a temperature of 0°C. Then, the middle $SiO_2$ was uniformly distributed by spin coating of commercially available Fox 16 resist at a speed of 5000 rpm without ramping up the process (Fig. S13e). As a result, the notch of the bottom disk could be filled while the top surface of $SiO_2$ is flat, which promises a fully embedded environment. After that, $CF_4$ gas-based RIE was performed to thin the middle $SiO_2$ layer to 60 (280) nm for the configurations in near-field (intermediate) regimes. The thickness of the final $SiO_2$ layer was confirmed by a thin film analyzer (FILMETRICS, F50). Subsequently, another 50 nm a-Si layer was deposited using ICP CVD to prepare for the top disk (Fig. S13f). The open structure was settled after well-aligned patterning and exactly $Cl_2$-based etching (Fig. S13g). To fabricate the embedded structure, another $SiO_2$ layer was covered on top of the open structure by spin-coating (Fig. S13i). Similarly, the bi-layer metasurface with displacement can be realized through the intentional shift of exposure position (Fig. S13h). The $FaPbI_3$ QDs were deposited uniformly by drop cast to realize unidirectional spin emission (Fig. S13j).

**Quantum Emitter Synthesis** The $FAPbI_3$ QD synthesis has been elaborated in detail in Ref.[50]. The supernatant of the synthesized $FAPbI_3$ solution was separated after centrifugation for 10 minutes. The supernatant was then filtered using a microfilter with a 0.2 μm pore size and diluted 1:1 with toluene for the final solution. The bi-layer metasurface was treated with oxygen plasma at room temperature for 10 min to make the surface more adhesive. Immediately after, $FAPbI_3$ QD solution was drop-cast onto the sample in a nitrogen-filled glove box at room temperature. The samples were stored in the dark in a nitrogen atmosphere at room temperature until further use.

**Back-Focal-Plane Reflectance Measurement** The photonic bands' mode dispersions were measured in a homebuilt Fourier plane setup (Fig. S14 in the Supporting Information SI-11). A white-light beam generated by a halogen lamp was right (left) circularly polarized using a linear polarizer (Thorlabs LPVISE100-A) and a quarter-wave plate (Thorlabs AQWP10M-580) where the axis of the linear polarizer at 45° (or-45°) to the fast axis of the quarter-wave plate. Circularly polarized light was focused on the sample surface with an area of 50×50 μm² using a 50 × objective lens with a numerical aperture (NA) of 0.6. The reflected light was collected and imaged by two lenses and measured by the spectrometer. The angle-resolved spectrum was measured using a monochromator slit as the line aperture in the Fourier plane. The measured raw reflectance spectrum was normalized to the background reflectance of a flat silver film.

**Photoluminescence (PL) Spectra Measurement** The optical setup for the characterization of light emission is depicted in Fig. S15 in the Supporting Information SI-11. A continuous-wave (CW) laser at 532 nm was used as the excitation light source (laser power of 125 mW). It first went through a dichroic mirror (FF552-Di02-25x36) to be reflected on the sample surface. After that, the emitted light from perovskite passes a longpass filter (Thorlabs FELH0650) to cut the effect of the laser source. The circular polarization of light was characterized by a quarter waveplate and a linear polarizer. All measurements were conducted at room temperature.

## Authors' Contributions

D.G., S.A., C.W.Q., Z.D., and L.W. conceived the project idea. C.W.Q., Z.D., and L.W. supervised the project. D.G., C.W.Q., and L.W. developed the theoretical background. D.G. performed numerical simulations. S.A. and Z.D. fabricated the samples. H.Y.L.L. provided support on the SEM image, and F.T. prepared the substrates. L.J.L. and Z.K.T. synthesized the quantum dots. S.A. performed the optical measurements. S.G., J.X., and Z.D. assisted with the optical measurement setup. D.G., S.A., C.W.Q., Z.D., and L.W. wrote the manuscript. All authors contributed to the data analysis and proofread the manuscript.

## Acknowledgments

This work was supported by the Singapore University of Technology and Design for the Start-Up Research Grant SRG SMT 2021 169 and Kickstarter Initiative SKI 2021-02-14, SKI 2021-04-12; and National Research Foundation Singapore via Grant No. NRF2021-QEP2-02-P03, NRF2021-QEP2-03-P09, NRF-CRP26-2021-0004, and NRF-CRP22-2019-0006. In addition, Z.D. would like to acknowledge the funding support from the Agency for Science, Technology, and Research (A*STAR) under its Career Development Award grant (Project No. C210112019), MTC IRG (Project No. M21K2c0116 and M22K2c0088), DELTA-Q 2.0 (Project No. C230917005), and National Research Foundation Singapore via Grant No. NRF-CRP30-2023-0003.